\documentstyle[times,epsfig]{jaa}

\title[GMRT detection of a new WAT radio source]
{GMRT detection of a new wide-angle tail (WAT) radio source associated with the galaxy PGC 1519010\\ }
\author[Kantharia, Das, Gopal-Krishna]{N. G. Kantharia$^1$, M. Das$^2$, 
Gopal-Krishna$^1$ \\
$^1$ National Centre for Radio Astrophysics,  Tata Institute of Fundamental Research, \\
Post Bag 3, Ganeshkhind, Pune University Campus, Pune-411007, India  \\
$^2$ Raman Research Institute, Sadashivnagar, Bangalore-560080, India \\}

\begin{document}
\maketitle

\begin{abstract}

We report the serendipitous detection of a Wide-Angle-Tail (WAT) radio
galaxy at 240 and 610 MHz, using the Giant Metrewave Radio Telescope 
(GMRT). This WAT is hosted by a cD galaxy PGC 1519010 whose
photometric redshift given in the SDSS DR6 catalog
is close to the spectroscopic redshifts (0.105, 0.106 and 0.107) of 
three galaxies found within $4'$ of the cD. Using the SDSS DR6 we 
have identified a total of 37 galaxies within $15'$ of the cD, 
whose photometric redshifts are between 0.08 and 
0.14. This strongly suggests that the cD is associated with a group 
of galaxies whose conspicuous feature is a north-south chain of galaxies (filament)
extending to at least 2.6 Mpc. The $ROSAT$ All-Sky Survey shows 
a faint, diffuse X-ray source in this direction, which probably marks 
the hot intra-cluster gas in the potential well of this group. 

We combine the radio structural information for this WAT with the
galaxy clustering in that region to check its overall consistency with the
models of WAT formation. The bending of the jet before and after its 
disruption forming the radio plume, are found to 
be correlated in this WAT, as seen from the contrasting morphological
patterns on the two sides of the core.  Probable constraints imposed
by this on the models of WAT formation are pointed out. We 
also briefly report on the other interesting radio sources found in the proximity 
of the WAT.  These include a highly asymmetric double radio source
and an ultra-steep spectrum
radio source for which no optical counterpart is detected in the SDSS.

\end{abstract}


\section{Introduction}

Wide-Angle Tail (WAT) are a subset of radio galaxies
near the Fanaroff-Riley \nocite{fanaroff} (1974) luminosity transition, which have 
been extensively discussed because of their exclusive association with 
cluster dominant (cD) galaxies and also because of the abrupt flaring of 
their jets after maintaining a well collimated flow to distances 
$\ge 20$ kpc from the core (e.g. Owen\& Rudnick \nocite{owen} 1976, O'Donoghue et al.
\nocite{donoghue1} 1993)
The jet disruption is sudden unlike FR I jets and the resulting plumes 
are often sharply bent. On the other side, although the jets are collimated
like FR II galaxies, they do not terminate in hot spots.  
Search for explanations of the WAT phenomenon 
began in the 1980s (e.g. Burns \nocite{burns}1981; Eilek et al. \nocite{eilek}1984; 
Leahy \nocite{leahy1}1984; O'Dea \& Owen \nocite{dea}1985; O'Donoghue et al. \nocite{donoghue}1990). 
Since neither the bending of the radio plumes occurs universally,
nor is the bending of jet essential for its flaring and plume
formation,  jet bending is currently regarded as a phenomenon 
independent from  jet flaring (Hardcastle \& Sakelliou\nocite{hardcastle1} 2004; 
Hardcastle, Sakelliou \& Worrall\nocite{hardcastle2} 2005). 

Since the host galaxies of WATs are the dominant members of their group/cluster
and hence located close to the bottom of the gravitational
potential well, they are not expected to have a 
large motion relative to the intra-cluster medium (ICM).  This  situation
is not conducive to large ram pressure that 
could bend the jet/plume leading to the WAT morphology.  An alternative mechanism, based
on numerical simulations has been investigated by Loken et al.\nocite{loken}(1995),
Hooda \& Wiita \nocite{hooda}(1996) and Burns et al. \nocite{burns1}(1994).  
In this picture, the sudden disruption of the jet
and bending of its plume can arise, if upon crossing a sharp
transition between the interstellar medium (ISM) of its host galaxy and the ICM, the jet 
encounters a transonic crosswind of the ICM resulting from cluster 
merger (an analytical treatment of the jet propagation through an 
ISM/ICM interface can be found in Gopal-Krishna \& Wiita \nocite{gopal}1987). 
Another alternate mechanism proposed in Loken et al. (1995) invokes a jet 
crossing an oblique shock formed due to colliding clumps in the cluster. 
Strong support for the merging cluster scenario comes from the 
detection of X-ray elongations which trace the merger axis and are found to be in 
the direction of the WAT radio tails (Gomez et al. \nocite{gomez}1997) 
and perpendicular to the initial jet direction (Burns et al. \nocite{burns1}1994).
Another mechanism, by Higgins, O'Brian \& Dunlop \nocite{higgins}(1999), 
associates the jet flaring with its encounter with a discrete 
clump in the ICM.  While all these mechanisms seem plausible, it is
intriguing that the deep Chandra imaging, which is available for the
best known WAT, namely 3C 465 (Hardcastle et al. \nocite{hardcastle2}2005), 
has failed to reveal any discontinuity
in the external medium at the locations where the jets flare.  Moreover, for this same
WAT, Jetha et al. \nocite{jetha}(2006) \& Hardcastle et al. \nocite{hardcastle2}(2005) have 
argued that, if the jet/plume is
extremely light relative to the external medium, the speed of the host galaxy required for
jet bending is only $\sim 100$ kms$^{-1}$, which is not implausibly high even for the
central galaxies of rich clusters (much higher speeds can occur for galaxies in
merging clusters).  
Thus, the bending of the jet/plume in 
WATs can possibly be explained. 

The circumstance of jet termination in WATs has been investigated
in several studies. 
Hardcastle \& Sakelliou  \nocite{hardcastle1}(2004) have shown that the distance between 
the host galaxy and the base of the plume inversely correlate with the
cluster richness as quantified in terms of the temperature of the 
ICM (also, Jetha et al. \nocite{jetha}2006). Since WATs reside 
at or close to gravitational centres of clusters and groups of galaxies, 
these can also be used as a signpost for the cluster or group of galaxies. 
Indeed, Blanton et al.  (\nocite{blanton}2000, \nocite{blanton1} 2001, \nocite{blanton2} 2003) have 
identified several clusters using the WAT sources detected in the FIRST survey 
(Becker et al. \nocite{becker}1995).

In this paper, we report the GMRT detection of a WAT associated with 
the galaxy PGC 1519010 (SDSS J113920.37+165206). The radio source
was noticed by us while imaging the radio continuum 
of the low surface brightness galaxy UGC 6614. 
This was interesting, given that the galaxy PGC 1519010 has itself been 
catalogued as a low surface brightness galaxy (U1-3) by O'Neil et al. \nocite{neil}(1997). 
In this paper, we discuss the nature of this radio source and 
its optical/X-ray environment. A Hubble constant of 70 kms$^{-1}$Mpc$^{-1}$ 
is assumed.  

\section{PGC 1519010 - the host galaxy}

\begin{figure}
\epsfig{file=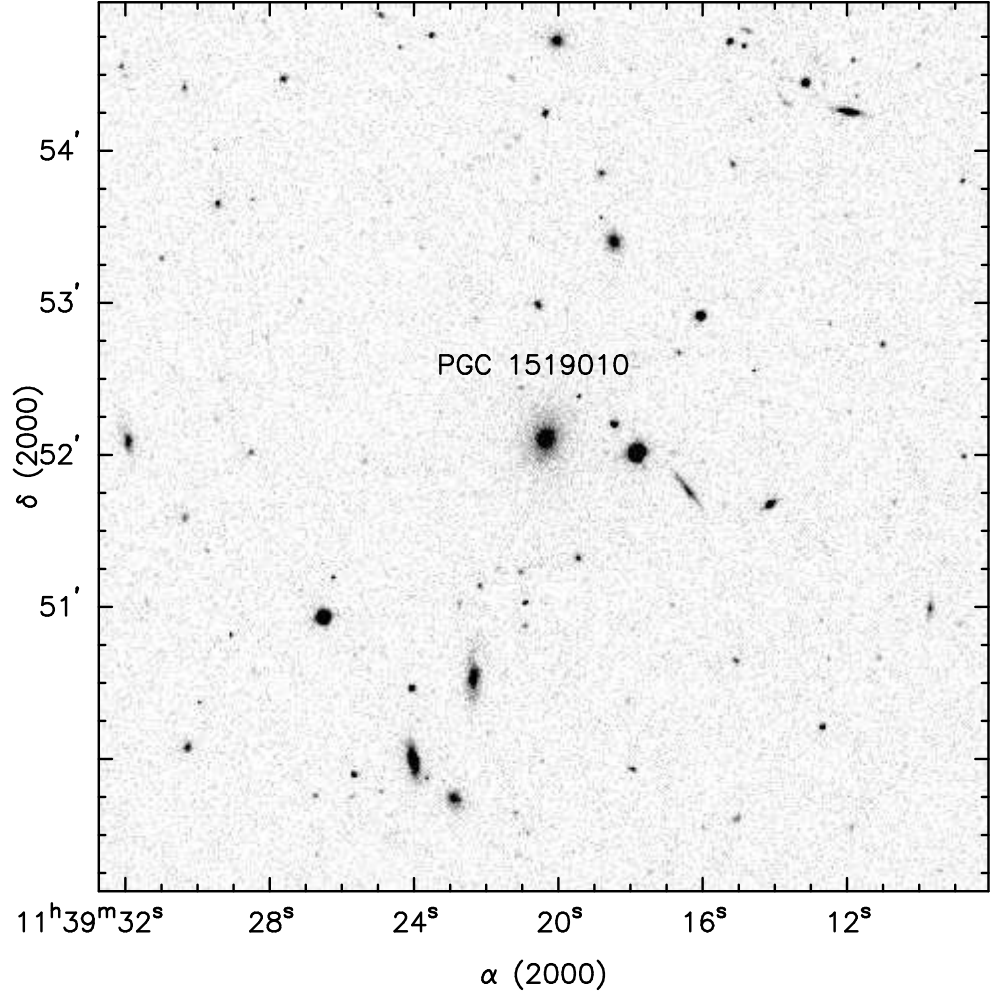,width=13cm}
\caption{PGC 1519010 (U1-3) in optical - from the SDSS.  
Notice the core-halo type of morphology 
indicating that it is a cD type elliptical with a large stellar halo extended north-south.  }
\label{fig1}
\end{figure}

PGC 1519010 is located $\sim16'$ to the south of UGC 6614, the giant 
low surface brightness (LSB) galaxy that was the principal target of our 
observations as part of our larger programme of observing the radio-band AGN
activity in giant low surface brightness galaxies 
(see Das et al. 2007\nocite{das}, 2008\nocite{das1}). 
We noticed that PGC 1519010 is catalogued by O'Neil et al. \nocite{neil}(1997) as the low
surface brightness galaxy U1-3.  The catalogue contains all those galaxies 
in the region of the Cancer and Pegasus clusters whose central surface brightness $\mu(0) 
\ge 22.0$ mag-arcsec$^{-2}$ (O'Neil et al. \nocite{neil} 1997). Curiously, O'Neil et al. 
found that U1-3 has a central surface brightness of $\mu_b(0) 
= 22.39$ mag-arcsec$^{-2}$ and that its radial brightness distribution is better 
fit by the King's model (1962\nocite{king}, 1966\nocite{king1}), rather than the exponential profile 
characteristic of LSB (disk) galaxies. Since King's model is known to 
describe the surface brightness distribution of globular clusters in our 
galaxy, O'Neil et al. had in fact suggested that U1-3 (PGC 1519010) might 
be a LSB globular cluster. A closer inspection of the SDSS image, 
however, showed that it is an elliptical galaxy, most probably a 
cD with a large halo extending in the north-south
(Fig. \ref{fig1}).  Such a core-halo distribution of stars can be
explained in terms of tidal distortion of an elliptical (with a typical
de Vaucouleurs r$^{1/4}$ brightness profile) by its repeated gravitational
encounters with other galaxies (e.g., Kormendy 1997), a highly plausible 
scenario for central regions of clusters and groups of galaxies.

Thus, we point out that the galaxy PGC 1519010 classified by O'Neil
et al. \nocite{neil} (1997) as an LSB galaxy needs to be
reclassified as an elliptical in view of its core-halo stellar
distribution (see Fig. \ref{fig1}), akin to cD galaxies generally found
near the centres of clusters/groups of galaxies. 
The main properties of this galaxy obtained from literature are summarized
in Table. \ref{tab1}.
\begin{figure}
\epsfig{file=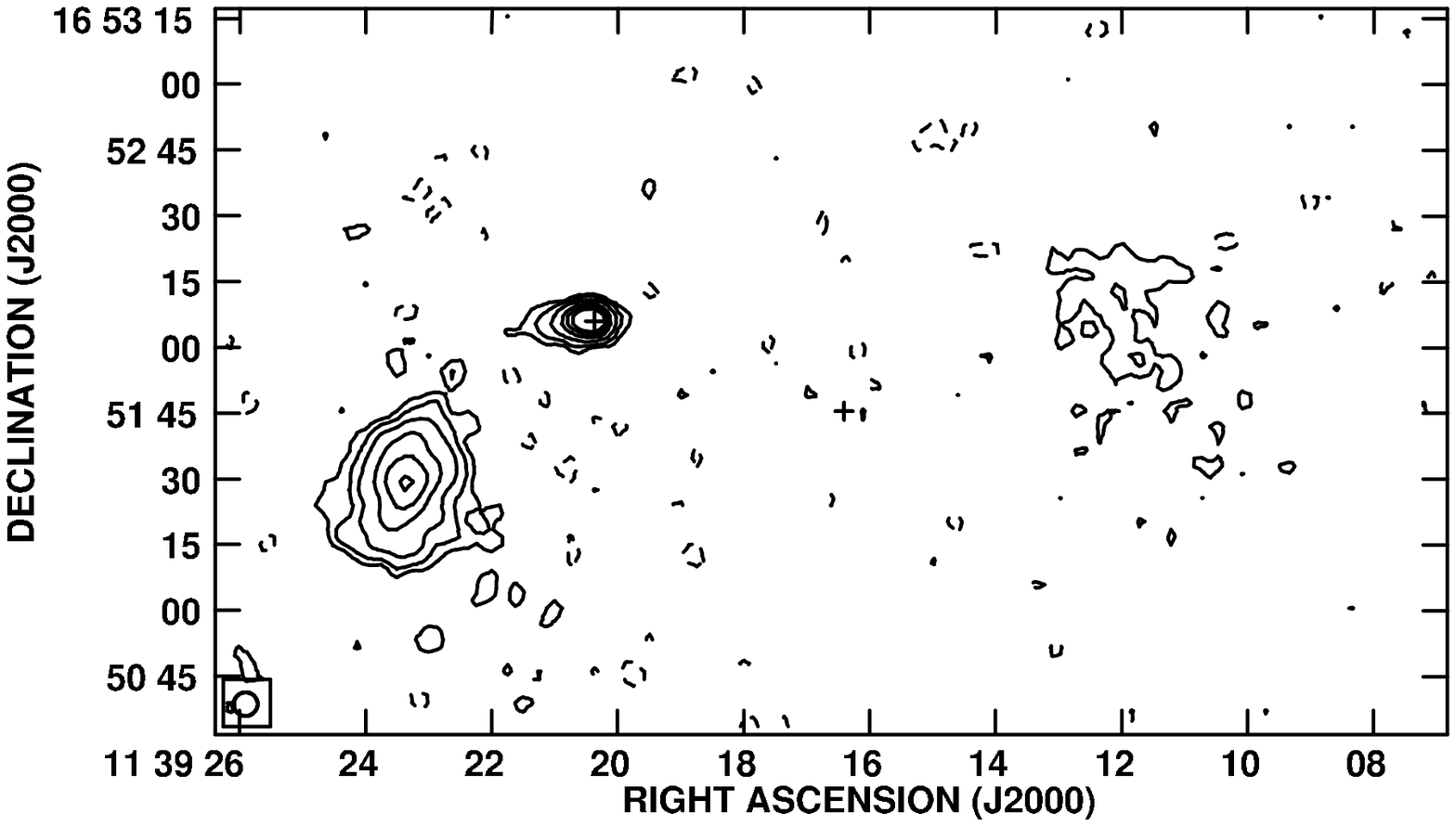,width=10cm}
\caption{(a) PGC 1519010 in 21cm radio continuum from the FIRST survey (resolution  $5.7''$ ).
Notice the double-lobed structure around the core of the galaxy.  
Notice the jets emerging from the central galaxy in the FIRST map.
The first contour is 0.36 mJy/beam
and thereafter the contours are plotted
in multiples of 2.  Notice the bent radio morphology on the eastern side of the core.  }
\label{fig2}
\end{figure}

\addtocounter{figure}{-1}
\begin{figure}
\epsfig{file=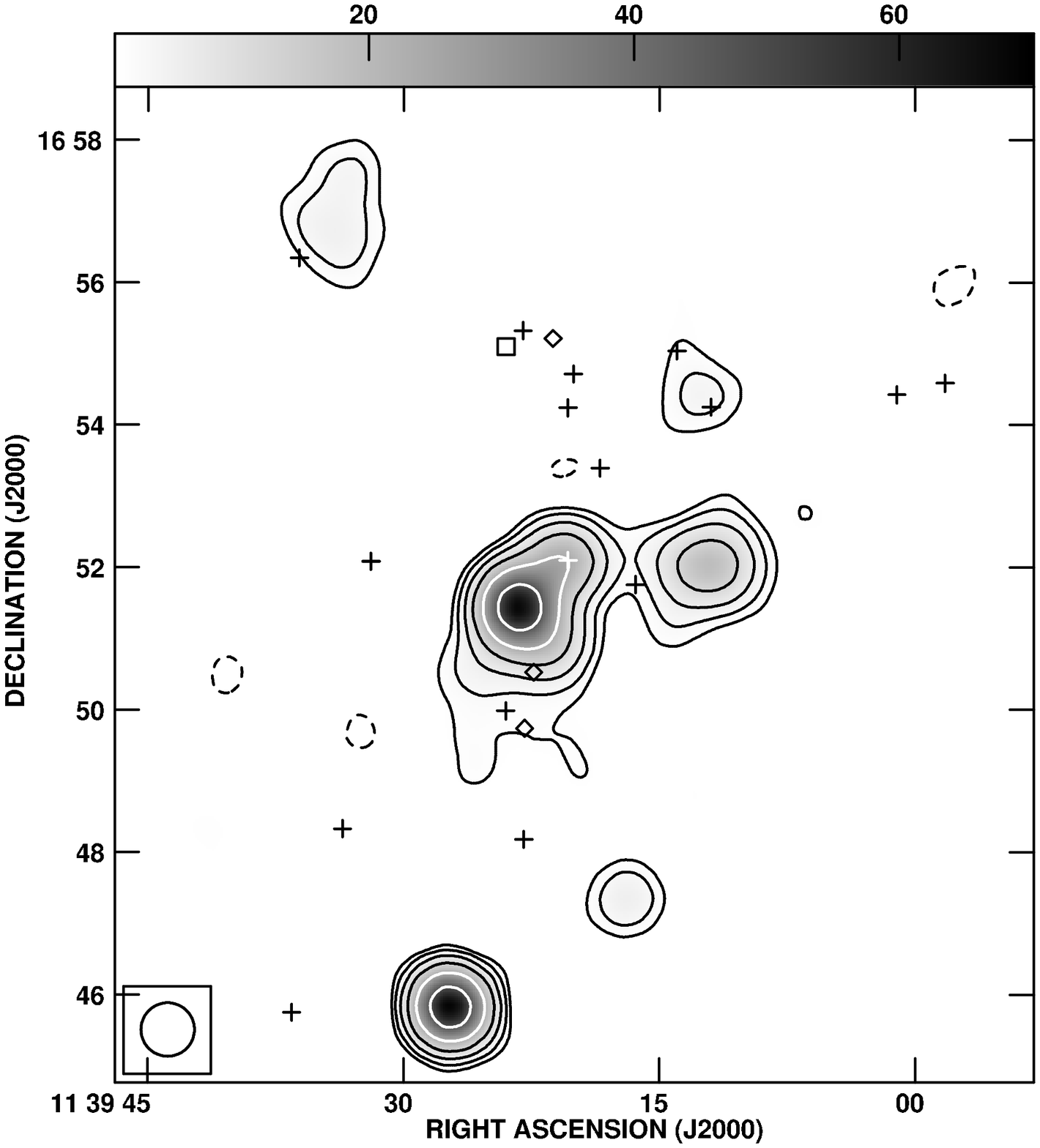,width=8cm}
\caption{(b) PGC 1519010 in 21cm radio continuum from the NVSS (resolution $45''$).
Notice the double-lobed structure around the core of the galaxy.
The first contour in the map is 1.5 mJy/beam and thereafter the contours are plotted
in multiples of 2.  Notice the bent morphology of the radio source. 
The plus (+) signs indicate the possible member galaxies of the group/cluster for which photometric redshifts
are available, the diamonds indicate member galaxies for which spectroscopic redshifts 
are available and the square indicates the centre of the cluster given by Gal et al. (2003)
which we discuss later in the paper.}
\label{fig2}
\end{figure}

Figs. \ref{fig2}a \& b shows the radio continuum images of the WAT radio galaxy at 
21 cm, reproduced from the FIRST survey (Becker et al. \nocite{becker}1995) and the 
NVSS survey (Condon et al. \nocite{condon}1998). 
The twin-lobed morphology consists of a diffuse radio lobe west of the 
core and a diffuse eastern lobe  which suggests that the eastern jet has undergone bending 
prior to flaring.  Such a morphology 
is indicative of WATs and we report here its structural details based 
on our GMRT observations. 
We also discuss the optical 
field around this galaxy, as well as the faint halo of soft X-rays 
picked up in the $ROSAT$ all sky survey (Voges et al. \nocite{voges}1999 and references therein).

\begin{table}[h]
\caption{Optical properties of PGC 1519010 from literature}
\begin{tabular}{l|c|c}
\hline
Parameter & Value & Reference \\
\hline
Names & SDSS J113920.37+165206 &  1 \\
      & PGC 1519010     & 2 \\
      & 2MASX J11392034+1652058  & 3 \\
Photmetric z1 & 0.138  & 1  \\
Photmetric zCC2 & 0.112  & 1\\
Photmetric zD1 & 0.104  & 1 \\
Radial brightness & Kings's model & 4 \\
Morphological type code & $2.1\pm5$ &  2 \\ 
Total apparent corrected B-mag & 16.69 &  2 \\
Total apparent corrected I-mag & 15.1 & 2  \\
Inclination & $35^{\circ}$ & 2  \\
u-magnitude & 18.72 & 1  \\
g-magnitude & 16.64 & 1  \\
r-magnitude & 15.62 & 1  \\
i-magnitude & 15.14 & 1  \\
z-magnitude & 14.83 & 1  \\
$\mu_b(0)$ mag-arcsec$^{-1}$ & 22.39 & 4 \\
Inclination angle & $47.9^{\circ}$ & 4 \\
Major axis upto $\mu_b=25$ isophote & $17.9''$ & 4 \\
Closeby cluster & NSC 113924+165506 & 5 \\
\hline \\
\end{tabular}
\\
$1.$ SDSS DR6 \\
$2.$ Hyperleda, Paturel et al. \nocite{paturel} (2003)  \\
$3.$ NED \\
$4.$ O'Neil et al. \nocite{neil} (1997) \\
$5.$ Gal et al. \nocite{gal} (2003) 

\label{tab1}
\end{table}

\begin{figure}
\epsfig{file=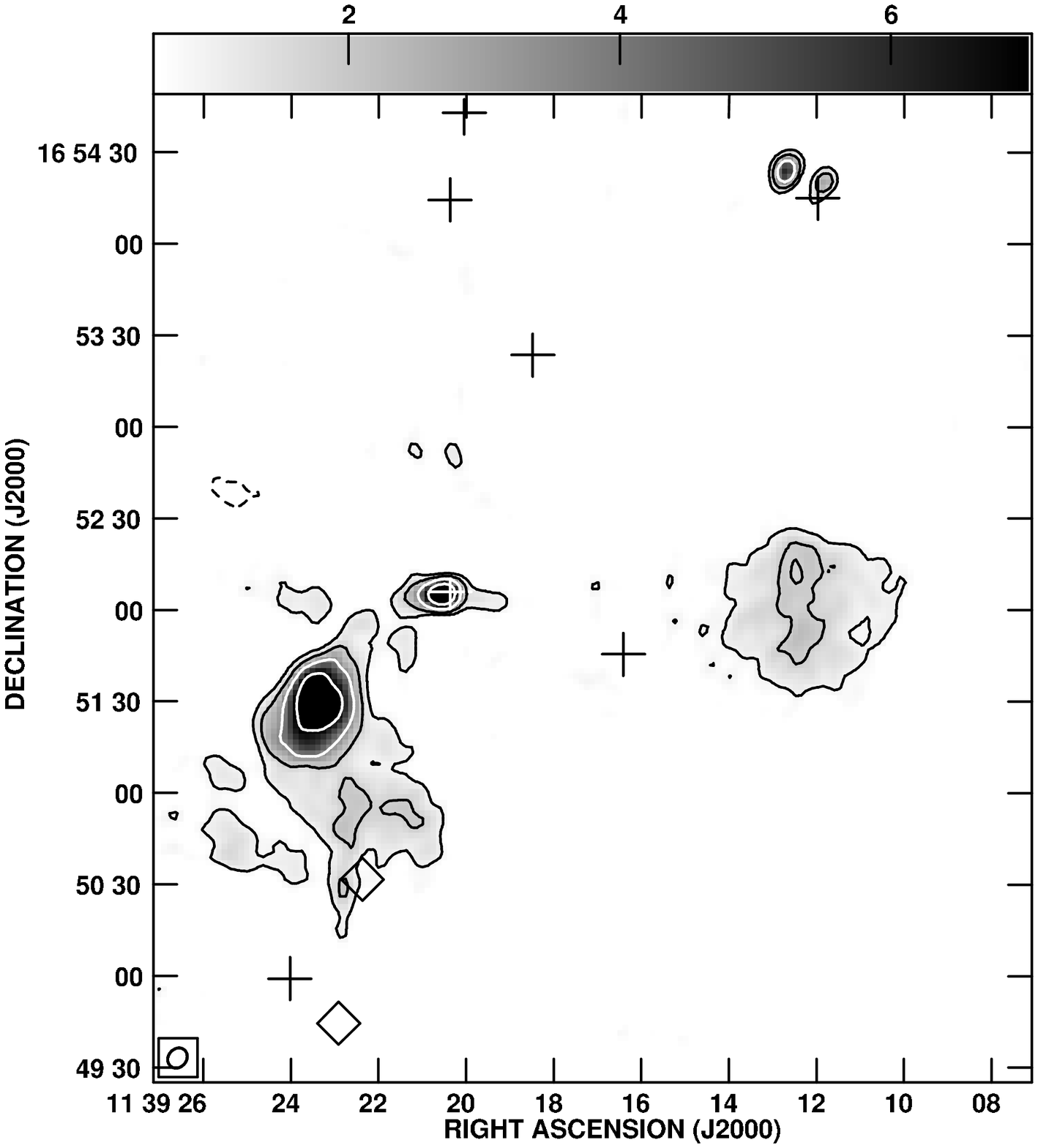,width=10cm}
\caption{(a) The WAT associated with PGC 1519010 in 610 MHz radio continuum imaged using the GMRT.  
This image was made using
Briggs robust=0 and has an angular resolution of $7''\times6''$ at a position angle
of $-40.3^{\circ}$.  The first contour is 0.9 mJy/beam and contours are subsequently plotted 
in multiples of 2.  Notice the  C-shaped morphology of the radio structure and the clearly bent
jet/plume in the east.  The explanation of the symbols are as in Fig. \ref{fig2}b.}
\label{fig3}
\end{figure}

\addtocounter{figure}{-1}
\begin{figure}
\epsfig{file=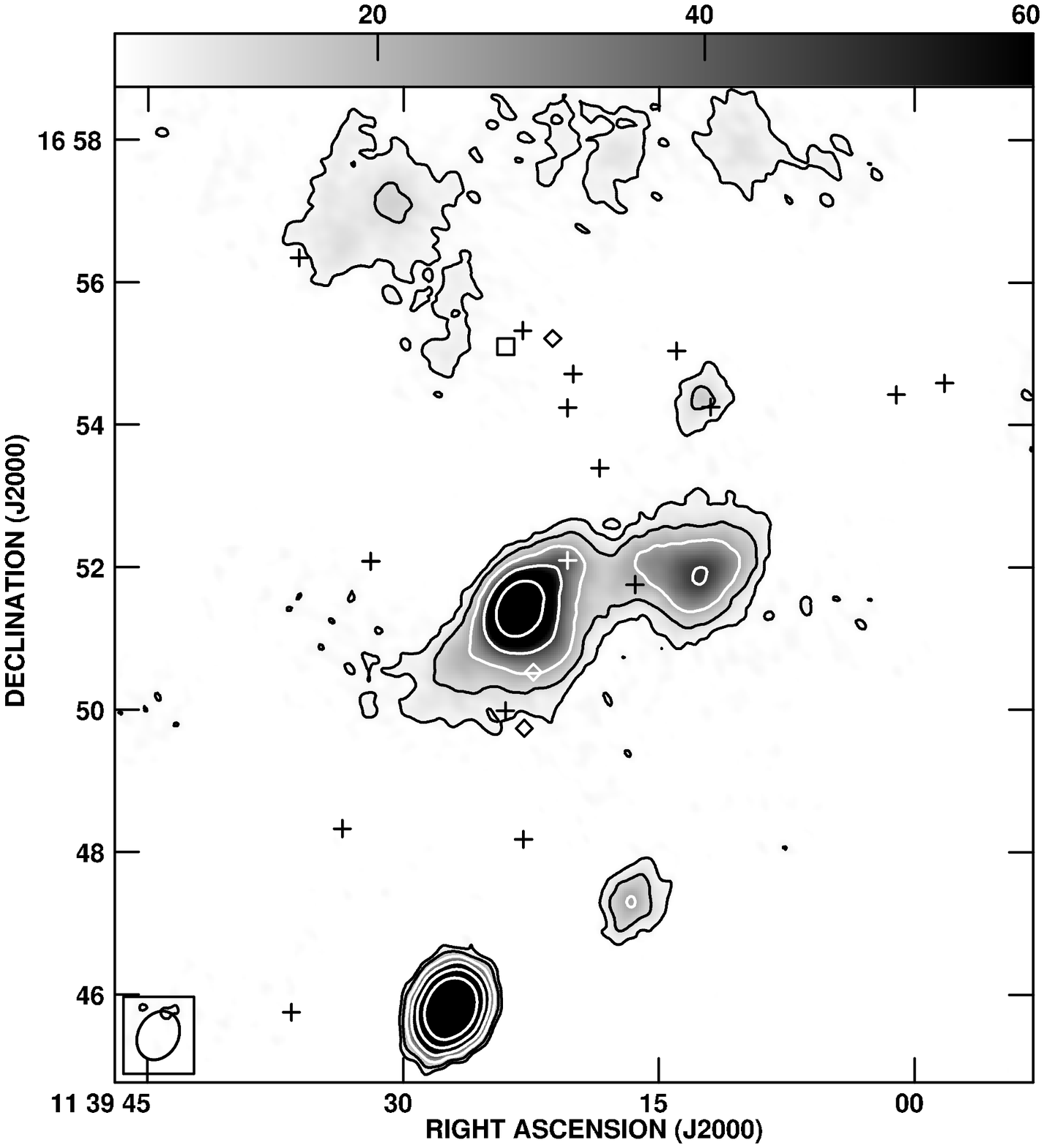,width=10cm}
\caption{(b) The WAT associated with PGC 1519010 in 240 MHz radio continuum imaged using the GMRT.
 Notice the double-lobed structure around the core of the galaxy.  
The 240 MHz image was made using natural
weighting with Briggs Robust=5 and has an angular resolution of about $43"\times34"$
at a position angle of $-29.4^{\circ}$.  The lowest plotted contour is 6 mJy/beam 
and the contours then increase in multiples of 2.  
The explanation of the symbols are as in Fig. \ref{fig2}b.}
\label{fig3}
\end{figure}

\begin{figure}
\epsfig{file=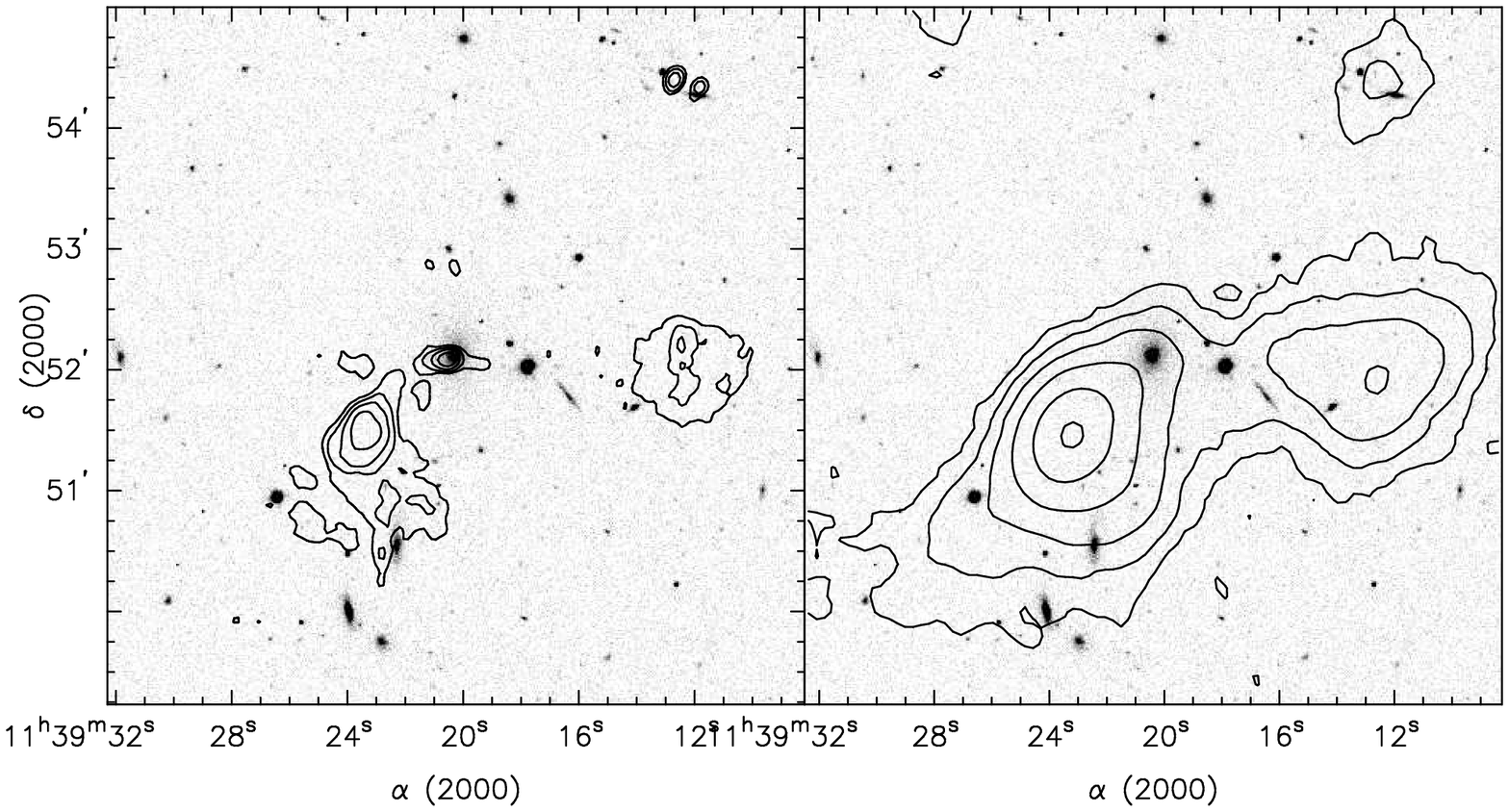,width=14cm}
\caption{The WAT radio source associated with PGC 1519010 in the 610 MHz (left) and
240 MHz (right) radio continuum superposed on the SDSS R band image plotted on the same angular scale.   
The angular resolution of the 610 MHz image is $7'' \times 6''$   
with a position angle of $-40.3^{\circ}$ whereas it is $43"\times34"$
at a position angle of $-29.4^{\circ}$ at 240 MHz.  The bright optical object between
the western jet and plume is classified as a star in the SDSS.  }
\label{fig4}
\end{figure}

\section{GMRT observations and results}
The field was observed on 30 December 2005 using GMRT (Swarup et al. \nocite{swarup}1991; 
Ananthakrishnan \& Rao \nocite{ananth}2002), in the dual-frequency mode
which allows simultaneous observations at 240 and 610 MHz (for details 
of the observations, see Das et al. 2008, in preparation). The data obtained in the 
native $lta$ format were imported to and analysed using 
NRAO AIPS\footnote{The National Radio Astronomy Observatory is a 
facility of the National Science Foundation operated under cooperative 
agreement by Associated Universities, Inc.}. 
Data of single RFI-free channels were first gain-calibrated. Bandpass 
calibration was then applied, after which several frequency channels 
were collapsed to generate a continuum database.  The 240 MHz data
were severely affected by radio frequency interference with the net 
result that only 1.2 MHz of the
total 6 MHz was usable and the rest of the data had to be discarded.  
The 610 MHz images were generated using a total bandwidth 
of 12.5 MHz.  We made images employing the robust weighting scheme 
(Briggs \nocite{briggs}1995), setting robust$=0$ (between uniform and natural weighting) 
and robust$=5$ (natural weighting) at both frequencies.  All the data
were self-calibrated and primary beam corrected.  
The GMRT images are shown in Figs. \ref{fig3}a \& b.  We note that the robust=0 image
at 240 MHz has a highly elliptical beam whereas the robust=5 image at 610 MHz
does not add any more information to the WAT structure and hence these are not
presented here.  Since the WAT is located about 16' south of the phase centre of
our observations, the sensitivity is compromised, especially at 610 MHz where
the half power width of the primary beam is about 50'.

\begin{table}
\caption{Two interesting radio sources found within $15'$ of the WAT radio galaxy.  
The 1.4 GHz flux density for GMRT1 is from FIRST and for GMRT2 is from NVSS data. 
Note that the radio position of GMRT1 and the position of the optical
counterpart of GMRT2 are listed here.  We use $S\propto \nu^{\alpha}$. }
\begin{tabular}{c|c|c|c|c|c|c}
\hline
Object & $\alpha_{2000}$ & $\delta_{2000}$ & S$_{1420MHz}$ & S$_{610MHz}$ & S$_{240MHz}$  & $\alpha^{240}_{1400}$  \\
 & hh mm ss.s & dd ' ''.'' & mJy  & mJy  & mJy  & \\
\hline
GMRT1  &  11 39 27.3  & 16 45 50.0 & 68.7 & 136 & 578 &    $-1.2$   \\
GMRT2  &  11 38 23.1 & 16 51 50.1 &  24.8  & 34.5 &   107.3  & $-0.82$     \\
\hline
\end{tabular}
\label{tab2}
\end{table}

A striking feature of this WAT, evident from all the maps
is that the overall bending of the radio structure is 
conspicuous only on the eastern side of the core (see Figs. \ref{fig2}, \ref{fig3}).
More clarity about the morphology emerges
from a joint inspection of the GMRT 
(610 MHz) and the FIRST (1.4 GHz) maps, both having a resolution of 
$\sim 6''$ (Figs. \ref{fig2}a \& \ref{fig3}a).  It is seen that the twin jets emerge from 
the core along the east-west direction and form
their lobes. Evidently, only the brighter eastern jet undergoes bending.
This occurs gradually along the jet and the plume  
followed by a sharp bending (towards
a position angle of about $220^{\circ}$) of the tail (see Fig. \ref{fig3}a).
In stark contrast, no sign of a bent morphology is evident on
the western side, despite the sudden flaring of the jet after propagating
for nearly 230 kpc from the core.

\begin{figure} 
\epsfig{file=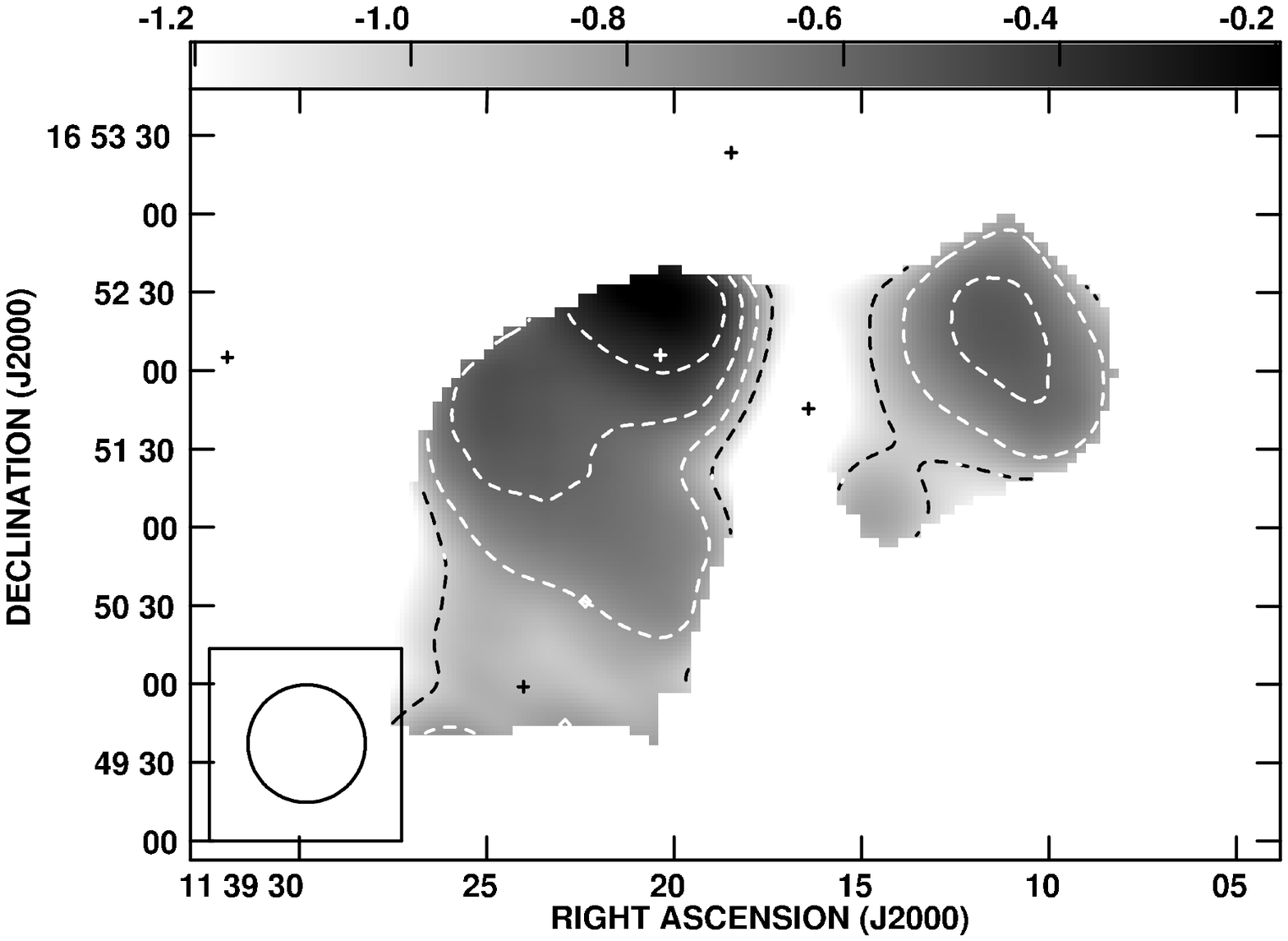,width=10cm} 
\%vspace{-2cm}
\caption{The low resolution ($45''$) spectral index distribution between 240 and 1.4 GHz.
Note the gradual steepening of the spectrum along the jet and plume in the east
with increasing separation from the core. } 
\label{fig5}
\end{figure}

In Fig. \ref{fig5} we show the spectral index distributions
derived by combining the GMRT images with the NVSS and FIRST images 
at 1.4 GHz (Fig.  \ref{fig2}).  For this, the GMRT 240 MHz map was first 
smoothed to the 45'' circular beam of the NVSS image.  To examine the spectral
index in the core region,
the spectral index map was generated by combining the GMRT 610 MHz and 
the 1.4 GHz FIRST maps, after convolving the FIRST image to match the
resolution of the 610 MHz GMRT image ($7''\times6''$). 
The spectral index distributions  were generated using the $>3\sigma$ emission
in the relevant maps.  The spectrum is flat near the 
core ($\alpha = -0.4)$ and starts steepening along the eastern jet.  The spectrum
further steepens to $\alpha$ $\sim -1 $ in the southernmost parts of the eastern
plume (see Fig. \ref{fig5}).   Near the peak of the eastern plume, we find $\alpha \sim  -0.7$.
Such a smooth variation in the spectral index 
away from the core is characteristic of WATs (see e.g., Hardcastle \nocite{hardcastle}1998).

\subsection{Interesting radio sources located near the WAT}

\begin{figure}
\mbox{
\epsfig{file=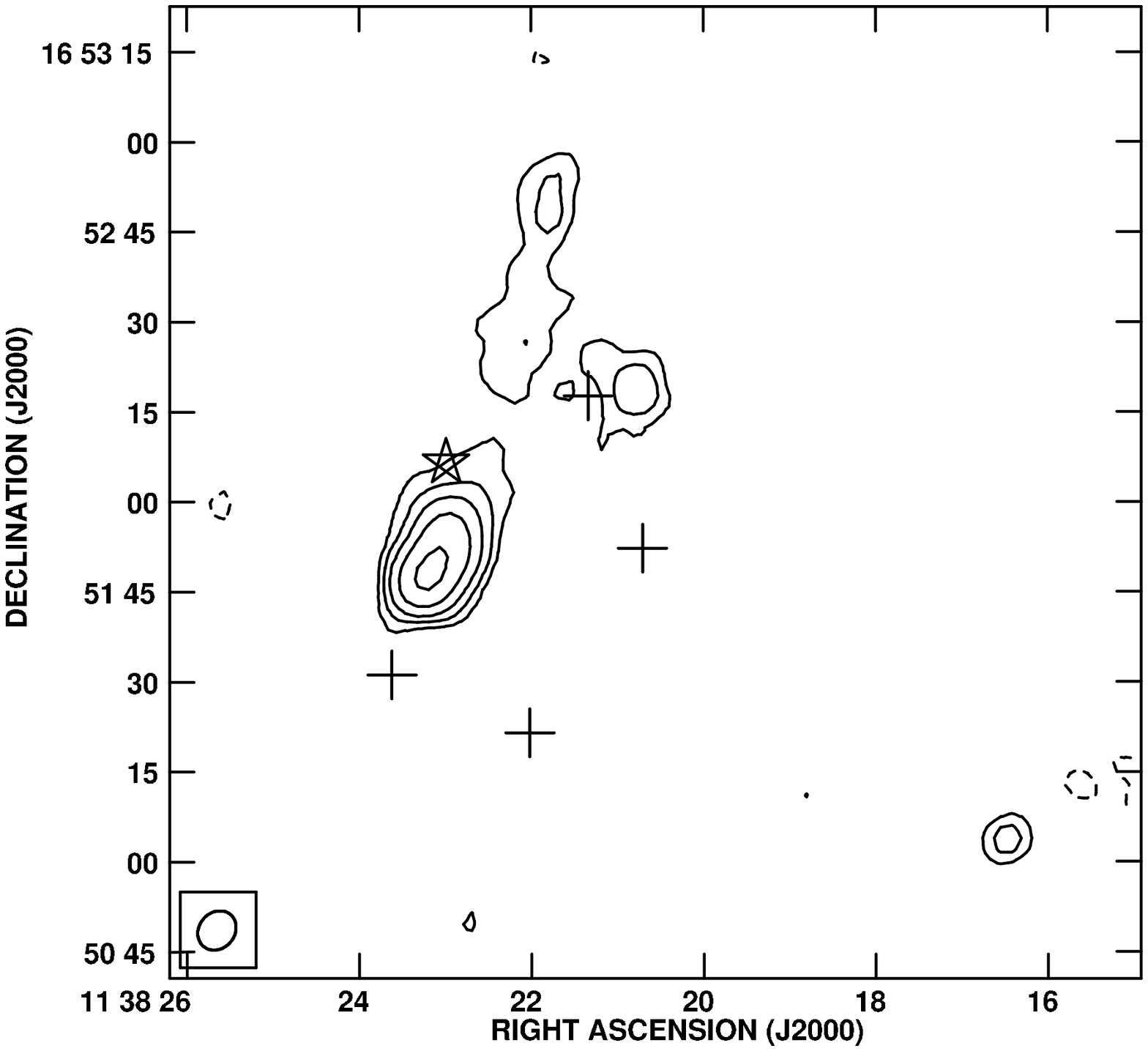,width=6.5cm}
\hspace{-0.5cm}
\epsfig{file=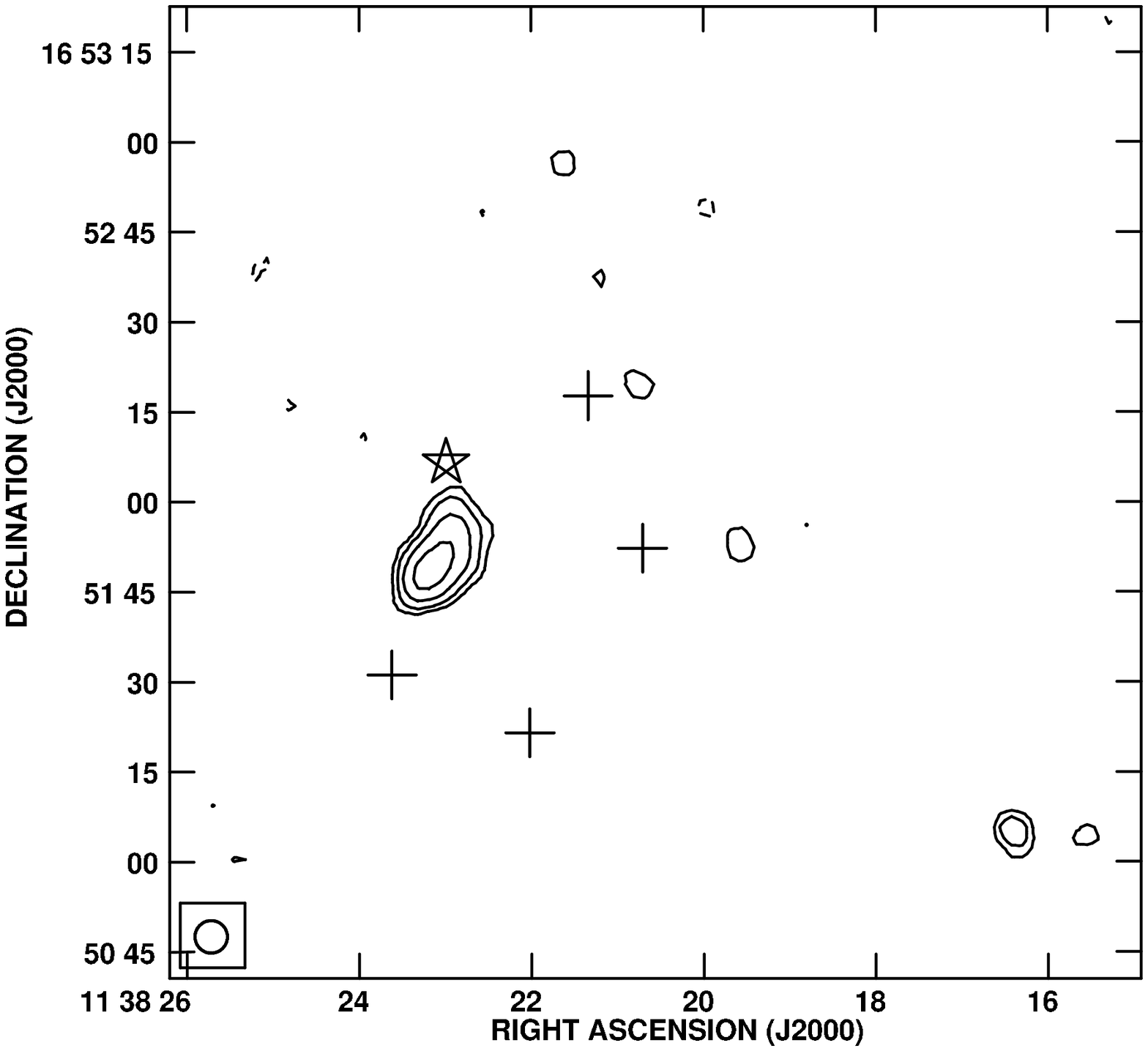,width=6.5cm}  }
\caption{The radio source GMRT2 located to the west of the WAT (Table \ref{tab2}). 
The 610 MHz GMRT image is shown on the left and the 1.4 GHz image from FIRST is shown
on the right.  The lowest contour at 610 MHz is 0.6 mJy/beam and at 1.4 GHz is 0.45 mJy/beam
and subsequently increase in multiples of 2.  
Notice the extended faint emission detected at 610 MHz which is not seen in the FIRST map.
Based on this detection, the star marks the position of the likely optical counterpart of the
asymmetric double radio source whereas the plus signs (+) mark the positions of the remaining
SDSS galaxies found within 30'' of the radio peak.}
\label{fig}
\end{figure}

We also determined the spectral indices of the radio sources found within half a degree 
of the cD galaxy.   The flux densities of the objects obtained
from our 240 MHz map and from the FIRST (Becker et al. \nocite{becker}1995) were
used unless the object was extended. (in which case NVSS data by 
Condon et al. \nocite{condon} (1998) were used).
Two interesting objects were found and we list their flux densities and spectral 
indices in Table \ref{tab2}. The first object (GMRT1) 
has an ultra-steep spectrum with $\alpha=-1.2$ and no optical counterpart is seen in the
SDSS.  The source is not resolved in the highest resolution maps which translates to an upper
limit on its angular extent of $1.8''\times0.6''$ at a position angle of $80^{\circ}$ obtained
from the FIRST.  The nearest galaxy listed by SDSS is $15''$ away from GMRT1.   This strong
ultra-steep spectrum object is located $\sim 3'$ to the south of the eastern lobe of the WAT
(Fig. \ref{fig2}b, \ref{fig3}b, \ref{fig7}).   

The second object (GMRT2) listed in Table \ref{tab2} lies to the west of the cD and
is an extended source.  SDSS lists five galaxies within $0.5'$ of the radio source
which are shown by crosses in Fig. \ref{fig}.  All are 
faint objects without reliable redshift estimates.   
Our 610 MHz image (Fig. \ref{fig}) has picked up faint extended emission which 
shows that the source is, in fact, a highly asymmetric double of total angular
size $\sim 80''$.  The likely counterpart is marked with a star in Fig. \ref{fig}. 
If this source lies at the distance of the WAT, then its linear size would be about 180 kpc. 
However since the host galaxy is faint, this object is likely to be more distant
and therefore physically large.

\section{Discussion}
\subsection{A new WAT and its associated cluster}

The galaxy PGC 1519010, as inferred here from its optical morphology and 
the association with a powerful WAT, is a cD galaxy (Table \ref{tab1}).

As mentioned above, the most striking and peculiar morphological feature 
of this WAT is the contrast between the jet/plume bending on the opposite
sides of the core, such that the bending is found exclusively on the
eastern side. In order to investigate this further, we have inspected
the SDSS database.  Gal et al. \nocite{gal}(2003), using the 
SDSS have identified a cluster (NSC 113924+165506) centred $\sim 3'$ 
north-east of PGC 1519010 (see Fig.\ref{fig3}, the square symbol) and having a photometric 
redshift of 0.1365, which is close to the value estimated for the WAT.
To probe this further, we show in Fig. \ref{fig7} 
the distribution of the galaxies listed in SDSS DR6.  For the cD progenitor
of the WAT, SDSS J113920.37+165206, the SDSS  provides three 
estimates of $z_{phot}$ = 0.138, 0.112 and 0.104. These are in reasonable 
agreement with the afore-mentioned value for the cluster ($z_{phot}$ = 
0.1365, Gal et. al. \nocite{gal}2003). In order to trace the optical field near this 
region we searched the SDSS for galaxies in the adjacent area and found 
three galaxies within $\sim 4'$ of the cD galaxy PGC 1519010 (WAT), having spectroscopic 
redshifts between 0.105 and 0.107.  These include U1-8 (U1-8, O'Neil et al. \nocite{neil}1997), 
a LSB galaxy lying about $2'$ south of the cD, for which Bergmann et al.
\nocite{bregmann}(2003) have estimated a spectroscopic redshift of 0.10715 (photometric redshifts 
given in the SDSS catalogue are 0.141, 0.095 and 0.091). The second galaxy, 
SDSS J113921.25+165512.8 is close to U1-8 and SDSS database gives a 
spectroscopic redshift of 0.105 whereas the photometric redshifts listed 
there are 0.121, 0.118 and 0.108. The third galaxy lies about $3.5'$ 
north of the cD and SDSS gives a spectroscopic redshift of 0.105 whereas 
the photometric redshifts are quoted there to be 0.061, 0.094 and 0.0.095.
All these spectroscopic redshift strongly suggest that the true redshift of 
the cD (WAT) is close to 0.106, which would also be in reasonable accord
with its photometric redshifts mentioned above. Taking $z$ = 0.106 for the 
cD, the radio luminosities of the WAT are 
$9.8\times10^{23}$ Watt-Hz$^{-1}$-Sr$^{-1}$ at 240 MHz and 
$3.2\times10^{23}$ Watt-Hz$^{-1}$-Sr$^{-1}$ at 1.4 GHz. 
These values lie in the region of the FR I/II transition and are thus  
characteristic of WATs.  
The distances from the core at which the jet flaring  
occurs are about 90 kpc and 230 kpc, for the eastern and western jets, respectively.  

To probe the large scale environment around the cD, we examined the 
SDSS DR6 data over a $1^{\circ}$ circular region centred at the cD. In particular, 
we searched for the galaxies for which a physical association with the 
cD is very probable. The criterion we employed was that the
photometric redshifts of the optical objects should be between 0.08 and 0.41.  A relatively
wide range had to be admitted since
the spectroscopic redshifts are available for only a few galaxies in the region.
We found that the range roughly represents the typical scatter among the three
SDSS estimates of the photometric redshifts for the objects in this region.  

The 37 objects that we selected from the above criterion include only five galaxies
with known spectroscopic redshifts.  Nonetheless, they can be expected to 
manifest the gross features of the galaxy clustering associated with 
the the cD (WAT) (see Fig. \ref{fig7}).  
Note also that in the soft X-ray band the
$ROSAT$ (Voges et al. \nocite{voges}1999 and references therein) 
database shows a diffuse source of about $0.5^{\circ}$ diameter, covering 
this region.  The detection of hot gas is consistent with the 
proposed galaxy clustering scenario.  As mentioned above, Gal et al. \nocite{gal}(2003) have
reported a galaxy cluster with Abell richness 0 which is close in both
redshift and direction to the galaxy group we identify here to be associated
with the WAT.

\subsection{Origin of the WAT}

\begin{figure}
\epsfig{file=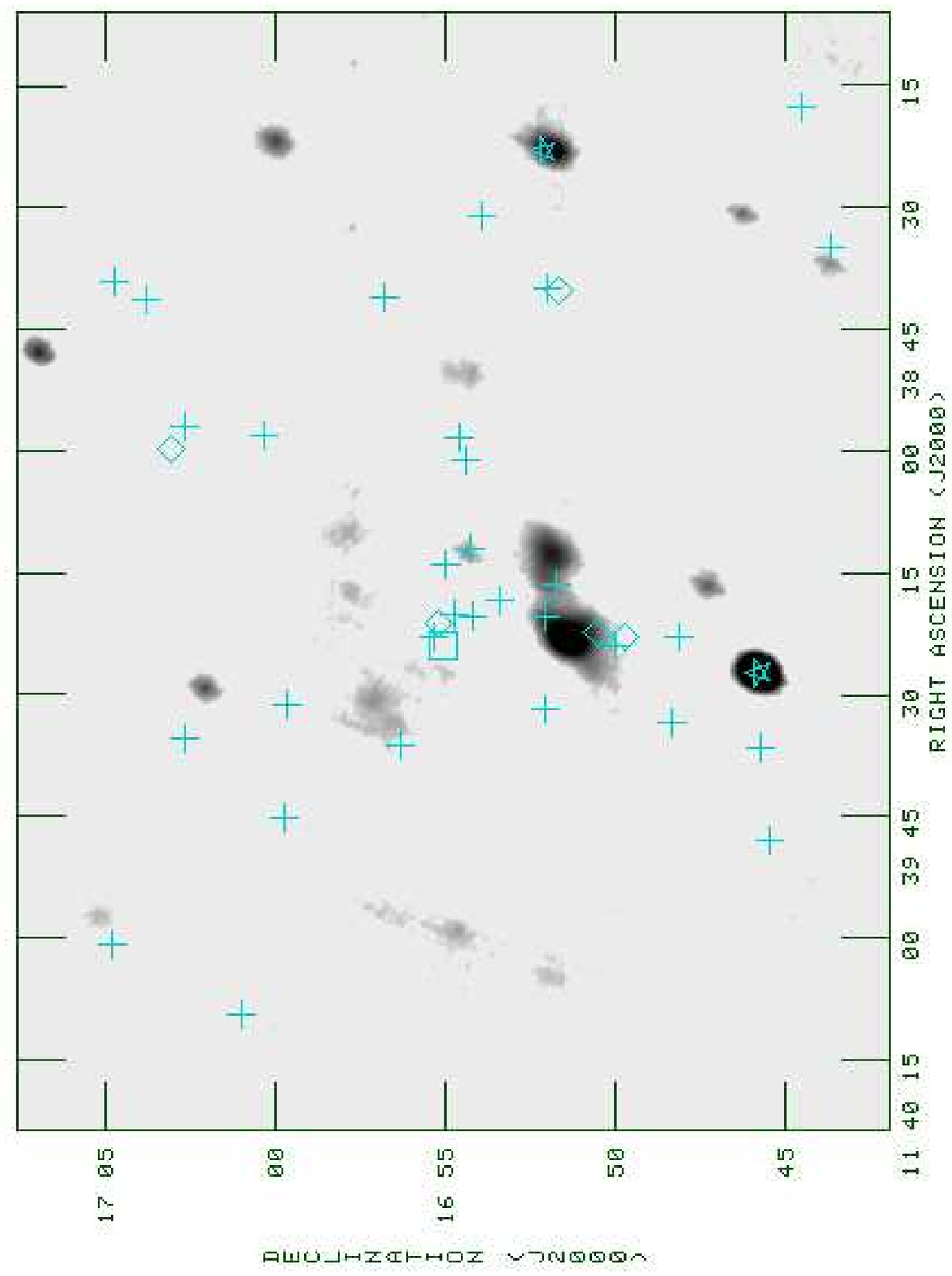,height=18cm}
\caption{240 MHz GMRT image of the WAT made with robust weighting=5 is shown.  
The positions of 37 galaxies which are likely to be associated with the WAT
are marked in the figure.  Five of them with known spectrocopic redshift are marked with diamonds.  
The remaining 32 member galaxies are marked by plus (+) sign.  The square marks
the position of the cluster centre NSC 113924+165506 ($z=0.1365$).
as given by Gal et al. (2003). The cluster is about 20'
($\sim 2.5 $ Mpc) in size.   Note the almost NS chain of galaxies extending southwards from the
cluster centre towards the cD.  The two interesting radio sources GMRT1 and GMRT2 (Table \ref{tab2})
are marked by stars.}
\label{fig7}
\end{figure}

Perhaps the most remarkable feature emerging from the galaxy distribution (Fig. \ref{fig7}) is
a nearly 0.5 Mpc long chain of galaxies stretching from the cD roughly 
north-south towards the cluster centre defined by Gal et al. \nocite{gal}(2003) and possibly
extending also to the south of the cD. It is 
along this filament that galaxies are likely to have approached the cD 
prior to merging with it.  This galaxy merger scenario is further supported by the 
shape of the stellar halo of the cD which too is extended roughly north-south. 
Thus it appears plausible that a bulk motion of the intergalactic gas has
been occuring along this galaxy filament. This circumstance may provide
potentially interesting clues about the mechanism of the jet disruption
and the jet/plume bending in this WAT.

Following upon the early attempts to understand the nature of WATs 
(e.g., Eilek et al. \nocite{eilek}1984, Leahy \nocite{leahy}1984, O'Dea \& Owen \nocite{dea}1985), 
a number of physical scenarios have been put forward to explain this 
rare type of phenomenon (Sect. 1).  In some of the models, the jet disruption 
occurs as the jet crosses the ISM of the host galaxy into the ICM, and is thereby 
subject to either a steep density gradient (Sakelliou \& Merrifield \nocite{sakelliou}1999), 
or a side-way ram pressure ("crosswind" arising from bulk motion 
of the ICM, see Loken et al. \nocite{loken}1995). In another scenario, the plumes 
form as the radio lobes of moderately powerful twin-jets are driven 
outwards due to buoyancy forces (Hardcastle \nocite{hardcastle3}1999). The latter model 
is motivated by the observations that the jet in some WATs continue 
well into the plume (e.g., Hardcastle \nocite{hardcastle3}1999; 
Hardcastle \& Sakelliou \nocite{hardcastle1}2004). 

Despite the modest sensitivity of our radio maps,  
the present WAT offers some insight and a broad check on some of 
the proposed models.  This is because of the morphological contrast 
observed between its two jet/plume structures, eventhough the kinetic powers 
of the two jets are expected to be similar.  Firstly, in this WAT a correlation is 
clearly seen between the bending properties of the jet and the resulting 
plume. As seen from Figs. \ref{fig2},  and \ref{fig4}, the western jet must be propagating straight 
for about 230 kpc before flaring and the resulting radio plume likewise 
shows no sign of bending. A contrasting pattern is seen on the eastern side 
of the core, where the collimated jet undergoes a steady bending 
until its disruption and thereafter the resulting plume too exhibits a 
sharp bend  consistent with the bent trajectory of the eastern jet.
From this correlated behaviour it appears that a viable mechanism for jet 
disruption in WATs should be able to bend both the plume and the associated jet. 
This requirement casts some doubt on the mechanism which seeks to 
explain the jet disruption in terms of collision with a dense gas cloud 
(Sect. 1).  Indeed, the observed alignment of the bent eastern 
plume with the chain of galaxies (see Fig. \ref{fig7}) would seem to be basically 
consistent with the ram pressure scenario involving bulk relative motion
between the ICM and the radio galaxy.  The observed radio structure
would also imply that the crosswind is in the NE-SW direction with the plume 
finally aligning with it (see Fig. \ref{fig3}a).
However, were the ram pressure of the ICM crosswind effective only after the jet crosses 
the ISM/ICM interface, it would be hard to explain the steady bending of 
the preceding collimated portion of the jet.  We may recall that a similar 
difficulty for the basic crosswind scenario (e.g., Loken et al. \nocite{loken} 1995) 
has been noticed in the case of the WAT 0647+693 where the 50 kpc 
long collimated western jet exhibits a steady bending before its disruption
(Hardcastle \& Sakelliou \nocite{hardcastle1} 2004).  It appears, therefore, that the effect of the 
ICM crosswind moving along the galaxy filament which may eventually cause
the sharp bending of the plume and its alignment along the filament also 
acts on the progenitor jet before its flaring. 
In fact, for the head-tail radio galaxy NGC 1265, Jones \& Owen \nocite{jones} (1979) have proposed
that a pressure gradient set up in the ISM due to the motion of the galaxy against
ICM can bend its jet within the ISM, provided their Mach number is not too large.
But, even if such a mechanism is viable for the present case, it will be required 
to explain the near absence of bending of the western jet and its radio plume (Fig. \ref{fig3}a \& \ref{fig4}). 
It would seem ad hoc to explain the contrasting 
pattern on the two sides by postulating a much larger kinetic power for the 
western jet. A conceivable alternative would be to postulate the existence of
inhomogeneities in the ICM on 100 kpc scale, but here too the
inhomogeneity will have to be positioned, so as to operate on 
just one side of the nucleus. Nonetheless, this possibility needs to be
investigated further. As a first step, it would be useful to make a deeper
radio image of this WAT, in order to look for any bent faint extensions of 
the western plume and to trace the trajectories of the two jets more clearly.
Secondly, this object is an excellent target for multi-object spectroscopy needed
for a better delineation of the galaxy distribution around the cD. Finally, 
targeted X-ray imaging of this region is needed to establish the 
morphological details of the $ROSAT$ detected diffuse X-ray source, so that 
the suspected filamentary gravitational potential well can be properly traced.

\section{Summary}
In this paper, we report the GMRT detection of a new wide-angle-tail radio source associated
PGC 1519010 which we find to be a cD galaxy.  These observations at 610 and 240 MHz were originally
aimed at studying the low surface brightness galaxy located 16' to the north.  
The power of the radio source at 240 MHz is $\sim 10^{24}$ Watt-Hz$^{-1}$-Sr$^{-1}$ which
is close to the FRI/FRII break.    Using the SDSS DR6 database, we have identified a cluster of 37 galaxies
likely to be associated with the WAT.  Further the ROSAT All Sky Survey shows
faint diffuse extended X-ray emission in the same region which indicates the presence
of a hot intracluster medium.   Using the spectroscopic redshifts available for a few galaxies
close to the cD, we suggest that the redshift of the group/cluster of galaxies associated
with the WAT is 0.106.  Moreover, we note that the clustering of member galaxies 
close to the WAT indicates a filament along which galaxies have been
merging with the cD.  This scenario also finds support in the observed NS extension of the
stellar halo of the cD galaxy. 

Our radio observations have highlighted the peculiar morphology of this WAT radio source. 
The source shows constrasting morphology on the two sides of the core: the eastern jet emerges
from the core and undergoes a gradual bending before being disrupted to form the plume
which itself continues to bend in the same direction.   On the western side,
the jet emerging from the core appears to continue straight for about 230 kpc, before
being disrupted to form the faint plume which too shows no bent extension.   
The very different behaviour on the two sides 
suggests that the mechanism responsible for the plume bending is also causing the jet to bend.
Several mechanisms for jet bending have been put forward and separate mechanisms
have also been invoked to explain the jet and plume bending (Sect. 1).  However, from our present results 
for the WAT associated with PGC 1519010, the similar bending shown by the jet and the plume
on the east of the core suggests that the same bending mechanism is operating on both components.

Several theories have tried to explain the WAT radio morphology.  From our data,
we favour the crosswind mechanism (Loken et al. \nocite{loken} 1995) wherein the jet 
runs into the crosswind due to the merging of clusters and which, in turn,
exerts ram pressure on the jet/plume, causing it to bend.  We note that the eastern jet continues for
about 90 kpc before being disrupted into a plume.  Around this distance, 
the jet would have crossed the hot ISM of the host elliptical galaxy. 
and would encounter the lower density ICM, causing it to flare.
The direction of the bending of the jet/plume agrees with the filament of galaxies and the extension
of the cD stellar halo.  Diffuse X-ray emission too is extended roughly in the NS direction, lending
further support to the crosswind mechanism. 
However, explaining the unaffected/unbent western jet/plume is difficult unless inhomogeneities in the
ICM are invoked.  More sensitive and high angular resolution observations are thus required to obtain
a better understanding of this interesting system.

We also report the detection of an ultra-steep spectrum source ($\alpha=-1.2$) to the south of
the eastern plume, which does not have an optical counterpart in SDSS.  
Another interesting source we have found is a highly asymmetric radio double source located to the west
of the WAT.

\section{Acknowledgements}
We thank the staff of the GMRT who have made these observations possible. 
GMRT is run by the National Centre for Radio Astrophysics of the Tata Institute of Fundamental Research.
This research has made use of the Sloan Digital Sky Survey (http://www.sdss.org)  Data Release 6 (SDSS DR6), 
the Faint Images of Radio Sky at Twenty cms (FIRST) survey, the NRAO VLA Sky Survey (NVSS),
NASA's Astrophysics Data System (ADS), ROSAT Data Archive (of the Max-Planck-Institut für 
extraterrestrische Physik (MPE) at Garching, Germany), 
NASA/IPAC Extragalactic Database (NED) (which is operated by the Jet Propulsion Laboratory, 
California Institute of Technology, under contract with the National Aeronautics and Space Administration), 
and Hyperleda (http://leda.univ-lyon1.fr).


\begin{thebibliography}{}
\bibitem{ananth} Ananthakrishnan, S. \& Rao, A. P., 2002, 
    \newblock{\it Multicolour Universe}, Ed. Manchanda, R. \& Paul, B., p233.
\bibitem{becker} Becker, R. H., White, R. L., Helfand, D. J., 1995, 
    \newblock{\it Astrophy. J.}, {\bf 450}, 559.
\bibitem{bergmann} Bergmann, M. P., Jorgensen, I., Hill, G. J., 2003,  
    \newblock{\it Astron. J.}, {\bf 125}, 116
\bibitem{blanton} Blanton, E.,  Gregg, M. D., Helfand, D. J., Becker, R. H., White, R. L., 2000, 
    \newblock{\it Astrophy. J.}, {\bf 531}, 118
\bibitem{blanton1} Blanton, E.,  Gregg, M. D., Helfand, D. J., Becker, R. H., Leighly, K. M., 2001, 
    \newblock{\it Astron. J.}, {\bf 121}, 2915
\bibitem{blanton2} Blanton, E.,  Gregg, M. D., Helfand, D. J., Becker, R. H., White, R. L., 2003, 
    \newblock{\it Astron. J.}, {\bf 125}, 1635
\bibitem{briggs}Briggs, D., 
    \newblock{\it High Fidelity Deconvolution of Moderately Resolved Sources},
PhD thesis, 1995
\bibitem{burns}Burns, J.O., White, R. A., Hough, D. H., 1981, 
    \newblock{\it Astron. J.}, {\bf 86}, 1
\bibitem{burns1}Burns, J. O., Rhee, G., Owen, F. N., Pinkney, J., 1994, 
    \newblock{\it Astrophy. J.}, {\bf 423}, 94 
\bibitem{condon}Condon, J. J., Cotton, W. D., Greisen, E. W., Yin, Q. F., Perley, R. A.,
Taylor, G. B., Broderick, J. J., 1998, \newblock{\it Astron. J.}, {\bf 115}, 1693 
\bibitem{das} Das, M., Kantharia, N. G., Ramya, S., Prabhu, T. P., McGaugh, S. S., Vogel, S. N.,
2007, \newblock{\it Mon. Not. R. Astron. Soc.}, {\bf 379}, 11
\bibitem{das1} Das, M., McGaugh, S. S, Kantharia, N. G.,  Vogel, S. N., 
 2008 in \newblock{\it Dark Galaxies and Lost Baryons}, Proceedings of the IAU Symposium, {\bf 244},  352. 
\bibitem{doe} Doe, S. M., Ledlow, M. J., Burns J. O., White, R. A., 1995, 
   \newblock{\it Astron. J.}, {\bf 110}, 46.
\bibitem{eilek} Eilek, J., Burns, J. O., O'Dea, C. P., Owen, F. N., 1984, 
   \newblock{\it Astrophy. J}, {\bf 278}, 37.
\bibitem{fanaroff} Fanaroff, B. L., Riley, J. M., 1974, 
   \newblock{\it Mon. Not. R. Astron. Soc.}, {\bf 167}, 31
\bibitem{gal} Gal, R. R., De Carvalho, R. R., Lopes, P. A. A., Djorgovski, S. G., 
Brunner, R. J., Mahabal, A., Odewahn, S. C., 2003, 
   \newblock{\it Astron. J.}, {\bf 125}, 2064
\bibitem{gopal} Gopal-Krishna, Wiita, P. J., 1987, 
    \newblock{\it Mon. Not. R. Astron. Soc.}, {\bf 226}, 531
\bibitem{gomez} Gomez, P. L., Ledlow, M. J., Burns, J. O., Pinkey, J., Hill, J. M., 1997,
    \newblock{\it Astron. J.}, {\bf 114}, 1711
\bibitem{hardcastle} Hardcastle, M. J., 1998, 
   \newblock{\it Mon. Not. R. Astron. Soc.}, {\bf 298}, 569
\bibitem{hardcastle3} Hardcastle, M. J.,  1999, 
   \newblock{\it Astron. \& Astrophy.}, {\bf 349}, 341
\bibitem{hardcastle1} Hardcastle, M. J., Sakelliou, I., 2004, 
   \newblock{\it Mon. Not. R. Astron. Soc.}, {\bf 349}, 560
\bibitem{hardcastle2} Hardcastle, M. J., Sakelliou, I., Worrall, D. M., 2005, 
   \newblock{\it Mon. Not. R. Astron. Soc.}, {\bf 359}, 1007
\bibitem{higgins} Higgins, S. W., O'Brien, T. J., Dunlop, J. S., 1999, 
   \newblock{\it Mon. Not. R. Astron. Soc.}, {\bf 209}, 273
\bibitem{hooda} Hooda, J. S., Wiita, P. J., 1996, 
   \newblock{\it Astrophy. J}, {\bf 470}, 21
\bibitem{jetha} Jetha, N.,  Hardcastle, M., Sakelliou, I., 2006, 
    \newblock{\it Mon. Not. R. Astron. Soc.}, {\bf 368}, 609
\bibitem{jones} Jones, T. W., Owen, F. N., 1979,  
    \newblock{\it Astrophy. J}, {\bf 234}, 818
\bibitem{king} King, I., 1962, 
    \newblock{\it Astron. J.}, {\bf 67}, 471
\bibitem{king} King, I., 1966, \newblock{\it Astron. J.}, {\bf 71}, 276 
\bibitem{kormendy} Kormendy, J., 1977, \newblock{\it Astrophy. J}, {\bf 218}, 333
\bibitem{leahy1} Leahy, J. P.,  1984, \newblock{\it Mon. Not. R. Astron. Soc.}, {\bf  208}, 323
\bibitem{leahy} Leahy, J. P., 1993,  Jets in Extragalactic Radio Sources, 
Proceedings of a Workshop Held at Ringberg Castle, Tegernsee, FRG, September 22-28, 1991. Ed:  H.-J. Röser 
and K. Meisenheimer. Springer-Verlag Berlin Heidelberg New York. 
Also Lecture Notes in Physics, volume 421, 1993, p.1
\bibitem{loken} Loken, C., Roettiger, K., Burns, J. O., Normal, M., 1995, 
    \newblock{\it Astrophy. J}, {\bf 445}, 80
\bibitem{neil} O' Neil, K., Bothun, G. D., Cornell, M. E., 1997, 
    \newblock{\it Astron. J.}, {\bf 113}, 1212
\bibitem{dea} O' Dea, C. P.,  Owen, F. N., 1985, 
    \newblock{\it Astron. J.}, {\bf 90}, 927 
\bibitem{donoghue} O'Donoghue, Aileen, A., Owen, F. N., Eilek, J. A.,  1990
    \newblock{\it Astrophy.J. Supple. Ser.}, {\bf 72}, 75
\bibitem{donoghue1} O'Donoghue, Aileen, A., Eilek, J. A., Owen, F. N.,  1993
    \newblock{\it Astrophy. J}, {\bf 408}, 428
\bibitem{owen} Owen, F. N., Rudnick, L., 
   \newblock{\it Astrophy. J}, {\bf 205}, L1
\bibitem{paturel} Paturel, G.,  Petit, C., Prugniel, P., Theureau, G., Rousseau, J., 
Brouty, M., Dubois, P., Cambr{\'e}sy, L.  2003, 
    \newblock{\it Astron. \& Astrophy.}, {\bf 412}, 45
\bibitem{sakelliou} Sakelliou, I., Merrifield, M. R., 1999, 
   \newblock{\it Mon. Not. R. Astron. Soc.}, {\bf 305}, 417
\bibitem{swarup} Swarup, G., Ananthakrishnan, S., Kapahi, V. K., et al. 1991,
    \newblock{\it Current Science}, {\bf 60}, 95.
\bibitem{voges} Voges, W., et al., 1999, 
    \newblock{\it Astron. \& Astrophy.}, {\bf 349}, 389
\end{thebibliography}
\end{document}